\documentclass[aps,pra,twocolumn,showpacs,floatfix]{revtex4}
\usepackage{epsfig}
\usepackage{graphicx}
\usepackage{dcolumn}
\usepackage{amsthm,amsmath}

\begin{document}
\title{Long-range interactions between the alkali-metal atoms and alkaline earth ions}
\author{Jasmeet Kaur$^a$, D. K. Nandy$^{b,c}$, Bindiya Arora$^a$ \footnote{Email: arorabindiya@gmail.com} 
and B. K. Sahoo $^{b,c}$\footnote{Email: bijaya@prl.res.in}}
\affiliation{$^a$Department of Physics, Guru Nanak Dev University, Amritsar, Punjab-143005, India}
\affiliation{$^b$Theoretical Physics Division, Physical Research Laboratory, Navrangpura, Ahmedabad-380009, India}
\affiliation{$^c$Indian Institute of Technology Gandhinagar, Ahmedabad, India}
\date{Received date; Accepted date}

\begin{abstract}
Accurate knowledge of interaction potentials among the alkali atoms and alkaline earth ions is very useful 
in the studies of cold atom physics. Here we carry out theoretical studies of the long-range interactions among 
the Li, Na, K, and Rb alkali atoms with the Ca$^+$, Ba$^+$, Sr$^+$, and Ra$^+$ alkaline earth ions systematically 
which are largely motivated by their importance in a number of applications. These interactions are expressed as 
a power series in the inverse of the internuclear separation $R$. Both the dispersion and induction components of 
these interactions are determined accurately from the algebraic coefficients corresponding to each power combination 
in the series. Ultimately, these coefficients are expressed in terms of the electric multipole polarizabilities of 
the above mentioned systems which are calculated using the matrix elements obtained from a relativistic coupled-cluster 
method and core contributions to these quantities from the random phase approximation. We also compare our estimated polarizabilities with 
the other available theoretical and experimental results to verify accuracies in our calculations. In addition, we also 
evaluate the lifetimes of the first two low-lying states of the ions using the above matrix elements. Graphical 
representation of the interaction potentials versus $R$ are given among all the considered atoms and ions.
\end{abstract}
\pacs{05.45.Ac,34.20.Cf,34.50.Cx}
\maketitle

\section{Introduction}\label{sec1}

Advancements in the simultaneous trapping and cooling of both ions and atoms in a hybrid trap \cite{Grier,Calarco} has in resulted 
significant upsurge in the precise description of the atom-ion interactions. This new development of using hybrid traps 
in which neutral atoms and ions are confined together lead to search for many exotic phenomenon in the quantum information 
science and condensed matter related fields~\cite{arne}. Interaction between these systems can be described as the 
special case of the van der Waal long range forces caused due to the fluctuating dipole moments of the systems \cite{dipole}. 
These interactions can enable many chemical reactions like charge-exchange and molecule formations at the single particle 
level, hence better understanding of these interactions is very useful in a number of studies such as explaining the underlying 
reasons for various quantum phase transitions \cite{saffman}, improvising quantum computing techniques~\cite{Joachim}, establishing 
sustained atom-ion sympathetic cooling mechanism~\cite{ravi,hall}, designing ultracold superchemistry~\cite{heinzen}, studying the 
physics of impurities in the Bose gases~\cite{goold,jogger}, interpreting cold atom collision processes~\cite{hall} etc.

Co-trapping of atoms and ions have several applications. Observations of the scattering between the atoms and the ions at low energy 
scale have been reported by a number of groups \cite{Rakshit,Ratschbacher,Willitsch}. Early studies on the properties of the mixed 
atom-ion systems were reported by C$\hat{\rm{o}}$t$\acute{\rm{e}}$ and his coworkers in order to investigate the ultracold atom-ion 
collision dynamics, charge transportation processes, and to realize possible formation of the combined stable system~\cite{cote1}.
Recently, H$\ddot{\rm{a}}$rter~\textit{et.al.} observed that the elastic scattering cross section of an atom-ion system depends on the 
collisional energy in the semi classical regime and favors scattering at small angles~\cite{arne}. Furthermore, the results of an 
atom-ion scattering event has been utilized to develop a novel and effective method to compensate excessive ion micro-motion in a 
trap~\cite{Huy}. Although there have been attempts to study the atom-ion interactions in the past, but the reported results
were not very accurate. Due to the experimental advancements in the atom-ion trapping experiments, it is now the time to 
provide more accurate description of these potentials to infer 
important signatures of new physics. Owing to the simplified and well understood structures of the alkali atoms and alkaline-earth ions,
they seem to be the natural choices and of immense interest for the experimental investigations \cite{sias} for which we intend to carry
out accurate theoretical studies of the long-range atom-ion interactions among these systems. In this work, we particularly undertake 
the Li, Na, K and Rb alkali atoms and the Ca$^+$, Sr$^+$, Ba$^+$ and Ra$^+$ alkaline earth ions to estimate their long range interactions.

Determination of the van der Waal coefficients of the atom-ion interactions require evaluation of the dynamic dipole and quadrupole 
polarizabilities at imaginary frequencies \cite{data}. We evaluate these polarizabilities by using dominant contributing 
matrix elements and experimental energies in a sum-over-states approach. These transition matrix elements are extracted 
either from the measurements of the lifetimes and the static dipole polarizabilities of the atomic states or using a relativistic coupled-cluster (RCC) 
method. Other contributions such as from the core and core-valence correlations, which cannot be estimated using the sum-over-states 
approach, are estimated using other suitable many-body methods. Unless stated otherwise, we use atomic unit (au) throughout this paper. 

\begin{table*}[t]
\caption{\label{lifetime}Contributions to the lifetimes of the $np_{1/2}$ and $np_{3/2}$ states of the alkaline 
Ca$^+$, Sr$^+$, Ba$^+$ and Ra$^+$ ions. The transitions rates ($A$s) are given in $10^6 s^{-1}$ and the lifetimes ($\tau$s)
are given in $ns$.}
\begin{ruledtabular}
\begin{tabular}{lccc|cccc}

 &  Ca$^+$ &  & & & Sr$^+$& & \\
$4p_{1/2}$ &  &  $4p_{3/2}$ & &  $5p_{1/2}$ &  &  $5p_{3/2}$& \\
\hline
 & & \\
$A$($4p_{1/2} \rightarrow 4s_{1/2}$)& 137.24 & $A$($4p_{3/2} \rightarrow 4s_{1/2}$)& 141.12 &  $A$($5p_{1/2} \rightarrow 5s_{1/2}$) & 130.1 & $A$($5p_{3/2} \rightarrow 5s_{1/2}$)       & 144.12 \\
$A$($4p_{1/2} \rightarrow 3d_{3/2}$)& 10.81  & $A$($4p_{3/2} \rightarrow 3d_{3/2}$) & 1.14   &  $A$($5p_{1/2} \rightarrow 4d_{3/2}$)& 9.22 & $A$($5p_{3/2} \rightarrow 4d_{3/2}$) & 1.17  \\
$\Sigma$ $A$                         & 148.05 & $A$($4p_{3/2} \rightarrow 3d_{5/2}$) & 10.17  &  $\Sigma$ $A$                        & 139.32& $A$($4p_{3/2} \rightarrow 3d_{5/2}$) & 9.89  \\
                                  &        & $\Sigma$ $A$       & 152.37 &                 &             & $\Sigma$ $A$              & 155.18 \\
$\tau(4p_{1/2})$                  &        & $\tau(4p_{3/2})$&        & $\tau(5p_{1/2})$&             & $\tau(5p_{3/2})$       &       \\
Present                           & 6.75                  &           & 6.55            & Present     & 7.16   &                        & 6.44 \\ 
Others                            & 6.88~\cite{safro}     &           & 6.69~\cite{safro}     & others  &  7.376~\cite{jiang}    & & 6.653~\cite{jiang}\\
Expt.                             & 6.96(35)~\cite{Ansbacher} &     & 6.71(25)~\cite{Ansbacher} & Expt.   &  7.35(30)~\cite{Gallagher} & & 6.53(20)~\cite{Gallagher}\\
                                   &                      &            &                      & Expt.   &  7.39(7)~\cite{Pinnington} & & 6.63(7)~\cite{Pinnington}\\
\hline
&  Ba$^+$ &  & & & Ra$^+$& & \\
 $6p_{1/2}$ &  &  $6p_{3/2}$ & &  $7p_{1/2}$ &  &  $7p_{3/2}$ &\\
\hline
& & \\
$A$($6p_{1/2} \rightarrow 6s_{1/2}$)& 95.13 & $A$($6p_{3/2} \rightarrow 6s_{1/2}$) & 119.88 &  $A$($7p_{1/2} \rightarrow 7s_{1/2}$)& 106.08 & $A$($7p_{3/2} \rightarrow 7s_{1/2}$)& 187.95 \\
$A$($6p_{1/2} \rightarrow 5d_{3/2}$)& 35.70 & $A$($6p_{3/2} \rightarrow 5d_{3/2}$) & 4.53 & $A$($7p_{1/2} \rightarrow 6d_{3/2}$)& 10.56 & $A$($7p_{3/2} \rightarrow 6d_{3/2}$) & 3.38 \\
$\Sigma$ $A$             & 130.83  & $A$($6p_{3/2}\rightarrow 5d_{5/2}$) & 35.30  &  $\Sigma$ $A$ & 116.64 & $A$($7p_{3/2} \rightarrow 6d_{5/2}$) & 22.89 \\
                      &         & $\Sigma$ $A$              & 159.72 &                        &        & $\Sigma$ $A$              & 214.23\\
 $\tau(6p_{1/2})$     &         & $\tau(6p_{3/2})$       &        & $\tau(7p_{1/2})$       &        & $\tau(7p_{3/2})$       &       \\
Present               & 7.64    &                        & 6.26   & Present                &  8.57  &                        & 4.66 \\ 
Others                & 7.83~\cite{Tchoukova}  &  & 6.27~\cite{Tchoukova} & Others         & 8.72~\cite{palR}&               & 4.73~\cite{palR}\\
Expt.                 & 7.74(40)~\cite{Gallagher}  &  & 6.27(25)~\cite{Gallagher} &     &                 &               &            \\
\end{tabular}
\end{ruledtabular}
\end{table*}

\section{Atom-ion interaction potentials}\label{sec2}

The long range potential $V(R)$ between an electrically charged ion and a neutral atom in their ground states, with $R$ as 
the internuclear distance, is divided in terms of the induced and dispersed interactions among the multipole moments as \cite{ca-mitroy,sandipan}
 \begin{equation}
 V(R)= V_{ind}(R) + V_{dis}(R),
 \label{VR}
 \end{equation}
 where $V_{ind}(R)$ and $V_{dis}(R)$ are known as the induced and dispersion potentials, respectively. It can be noted that a small 
contribution coming from the exchange potential \cite{sandipan} has been neglected in the above expression. The induced part of this 
potential occurs due to polarization from the attractive interaction of the permanent multipole of the ion with the induced multipole 
of the atom due to the ion and is expressed in terms of the induction coefficients ($C_{2n}$) as \cite{sandipan,ca-mitroy}
\begin{equation}
V_{ind}(R)= -Q^2\sum_{n=1}^{\infty} C_{2n}/R^{2n},\\
\label{c2n}\\
\end{equation} 
where $Q$ is the charge of the ion and negative sign indicates that the force is attractive in nature. In the present 
article, we have truncated the series at powers of $R^{-6}$ and contributions from the higher order coefficients associated 
with $R^{-8}$ and $R^{-10}$ terms are suppressed here. In the above equation, the term $R^{-2}$ which corresponds to the 
charge-dipole interaction vanishes for the interaction of an ion with a neutral atom. The second term inside the summation, 
corresponding to $n=2$, is a spherically symmetric term arising due to the ion-induced dipole potential and is given as $C_4/{R^4}$ 
with $C_4=\alpha_1/2$ for the static dipole polarizability $\alpha_1$ of the atom. This term originates due to the electric 
field created by the ion which induces an electric dipole moment in the neutral atom. This part of the potential is independent 
of the electronic state of the ion, but varies with the electronic state of the atom due to the dependencies on their $\alpha_{1}$s.
Once the $C_4$ coefficients are known, one can also calculate the characteristic length scale ($R^*$), the effective range of the 
polarization potential, by equating the potential to the kinetic energy as ${R^*} $=$\sqrt{2\mu{C_4}}$ \cite{arne,doerk}. The characteristic 
energy scale is further expressed in terms of $R^*$ as $E^*$=1/$2\mu {R^*}^2$. Here $\mu$=($m_{ion}$)($m_{at}$)/($m_{ion}$+$m_{at}$)
is the reduced mass of the system for the mass of the ion $m_{ion}$ and mass of the atom $m_{at}$. The next term with powers of $R^{-6}$ in 
the general expression (Eq.(\ref{c2n})) appears due to the instantaneous fluctuating dipole moments between the atoms and can be expressed 
as $\frac{C_6}{R^6}$ with $C_6$ =$\alpha_2$/2 for the quadrupole polarizability $\alpha_2$ of the atom.

 For the atom and ion being in their respective ground states, the expression for the dispersion interaction potential is given by
 ~\cite{cote1,makarov}
\begin{equation}
 V_{dis}(R)=-\frac{c_6}{R^6} - \frac{c_8}{R^8} -\frac{c_{10}}{R^{10}}\cdots
\label{eqdis}\\
\end{equation}
The coefficients $c_6$, $c_8$, $c_{10}$, $\cdots$ etc. emerge from the instantaneous dipole-dipole, dipole-quadrupole, dipole-octupole, 
quadrupole-quadrupole, etc. interactions and are known as the dispersion coefficients. In the long-ranged potential, first 
term dominates over the other terms and the higher order terms are sufficiently weak to be neglected. In Eq. (\ref{eqdis}), the 
dispersion coefficient $c_6^{AB}$ between an atom $A$ and an ion $B$ can be estimated using the expression given as~\cite{Koutseiosa} 
\begin{equation}
c_6^{AB} =\frac{3}{\pi}\int_0^{\infty}d\omega \alpha_1^{A}(\iota\omega)\alpha_1^{B}(\iota\omega).
\label{eqc6}\\
\end{equation} 
Here $\alpha_1^{A}(\iota\omega)$ and $\alpha_1^{B}(\iota\omega)$ are the atomic and ionic polarizabilities respectively. Since it is
cumbersome to determine these dynamic polarizabilities for a sufficiently large number of frequencies, therefore 
instead of using the exact {\it ab initio} methods alternative approaches have been adopted to calculate the $c_6^{AB}$ 
coefficients in the literature. Among these the Slater-Kirkwood formula~\cite{sk} is one of the popular methods~\cite{Koutselos2} in which the 
dispersion coefficients for atom-ion system are approximated by
\begin{equation}
c_{6}^{AB}={\frac{3}{2}}{\frac{\alpha_1^{A}\alpha_1^{B}}{(\alpha_1^{A}/N^{A})^{1/2}+(\alpha_1^{B}/N^{B})^{1/2}}},\label{slater}
\end{equation}
where $N^{A}$ and $N^{B}$ are the effective number of electrons and determined using the following empirical formula which assumes that
the dominant contributions arise from the loosely bound electrons present in the outer shell of the systems
\begin{equation}
\label{number}\\
(N^{A})^{1/2}=\frac{4}{3}{c_6^{AA}}/{(\alpha_1^{A})^{3/2}},
\end{equation}
with the van der Waals coefficient $c_6^{AA}$ of the homo-nuclear dimer and static polarizability $\alpha_1^A$ of the atom $A$.
This approximation may work reasonably if the dynamic polarizabilities are very large for lower frequencies, falling 
swiftly towards the asymptotic region of the frequencies and when the trends of the dynamic polarizabilities are almost 
same in both the coupled atomic systems. Substituting the above relation, we get 
\begin{equation}
\label{modified}
c_{6}^{AB} =\frac{2c_6^{AA}c_6^{BB}}{\alpha_1^{B}\alpha_1^{A}c_6^{AA}+\alpha_1^{A}\alpha_1^{B}c_6^{BB}}.
\end{equation}
Another approximation to calculate the dispersion coefficients among the hetero-nuclear alkali dimers has been considered by Derevianko
~\textit{et. al.}~\cite{babb} as 
\begin{equation}
 c_6^{AB}=\frac{1}{2}{\sqrt{c_6^{AA}c_6^{BB}}}\frac{\Delta{E^{A}}+\Delta{E^{B}}}{\sqrt{\Delta{E^{A}}.\Delta{E^{B}}}}. 
 \label{aa}
\end{equation}
In this approach, it is assumed that the most contribution to $c_{6}$ coefficient comes from a principal transition in each
system whose transition energies are denoted by $\Delta{E^{A}}$ and $\Delta{E^{B}}$. Nevertheless, both the above approximations are only
valid for the qualitative description of the atom-ion interaction potentials, but it is imperative to use more accurate values
of the dynamic the multipole polarizabilities for the precise description of the atom-ion interaction potentials. In our earlier 
works, we had determined dynamic dipole polarizabilities of the alkali atoms for a sufficiently large number of imaginary frequencies 
very precisely \cite{arora,bk}. In the present work, we determine further these quantities for the alkaline earth ions and quadrupole
polarizabilities of the alkali atoms in order to determine the above discussed van der Waals coefficients accurately. We compare
these coefficients with the values obtained using the Slater-Kirkwood formula~\cite{sk} given by Eq. (\ref{eqc6}) and with the 
approximation used by Derevianko and coworkers~\cite{babb} in the previous studies. Moreover, we also determine the lifetimes of the 
first excited $np$ states of the alkaline earth ions and compare them with the available experimental and other precise calculations 
in order to test the accuracies of the dipole matrix elements of the transitions that are predominantly contributing in 
the determination of the dipole polarizabilities of the considered ions.

\begin{table*}[t]
\caption{\label{pol11} Calculated values of the static dipole and quadrupole polarizabilities along with the $C_4$ and 
$C_6$ coefficients for the Li, Na, K, and Rb alkali atoms. Polarizability values are compared with the other available 
theoretical and experimental results. References are given in the square brackets.}
\begin{ruledtabular}
\begin{tabular}{lcccc}
Polarizabilities & Li & Na & K & Rb \\
\hline
 & & \\
$\alpha^{vm}_1$              & 162.5    & 161.9   & 284.3    &  309.1  \\
$\alpha^{c}_1$                & 0.2      & 0.9     & 5.5      & 9.1  \\
$\alpha^{vc}_1$              & 0.0      & 0.0     &-0.1      & -0.3    \\
$\alpha^{vt}_1$              & 1.2      & 0.08    & 0.06     & 0.11    \\
$\alpha^{total}_1$ (Present) & 164.1(6) & 162.4(2)& 289.8(6) & 318.3(6)\\
$\alpha^{total}_1$ (Other)      & 164.112(1)~\cite{Tang}   & 162.9(6)~\cite{Thakkar} & 289.3~\cite{safronova}       & 315.7~\cite{mitroyrb}\\
$\alpha^{total}_1$ (Expt.) & 164.2(1.1)~\cite{miffre} & 162.1(8)~\cite{ek}      & 290.58(1.42)~\cite{holmgren} &  318.79(1.42)~\cite{holmgren}\\ 
$C_4$ coefficient               & 82.1 & 81.2 & 144.8 & 159.9\\
\hline
& & \\
$\alpha^{vm}_2$              &1345    &1780   &4839    &6244 \\
$\alpha^{c}_2$               & $\sim 0$      & 2     &16      &35   \\
$\alpha^{vc}_2$              &0       &0      &0      &0   \\
$\alpha^{tail}_2$            &81      &113    &94      &211  \\
$\alpha^{total}_2$(Present)  &1426    &1895   &4947    &6491  \\
$\alpha^{total}_2$(Other)    &1424~\cite{qporsev} & 1879~\cite{spelsberg}, 1902~\cite{makarov} & 5000~\cite{Marinescu}  & 6459~\cite{Marinescu}\\
$C_6$ coefficients          & 713    & 947   &  2474 &  3245\\
\end{tabular} 
\end{ruledtabular}
\end{table*}

\section{Evaluation of multipolar polarizabilities}\label{sec3}

The dynamic dipole (E1) and quadrupole (E2) polarizabilities of the atomic systems with an imaginary frequency $\iota \omega$ are 
given by
\begin{equation}
 \alpha_k(\iota \omega) =- \sum_{I \ne n} \frac{ (E_n - E_I) |\langle \Psi_v | O_k | \Psi_I \rangle |^2}{(E_n - E_I)^2 + \omega^2},
 \label{poleq}
\end{equation}
where $n$ is the principal quantum number of the ground state of the respective system, $I$ represents all possible 
allowed intermediate states, $k=1$ and $O_1 \equiv D=|e|r$ for the dipole 
polarizability ($\alpha_1$) and $k=2$ and $O_2 \equiv Q=\frac{|e|}{2}(3z^2-r^2)$ for the quadrupole polarizability ($\alpha_2$). For 
the {\it ab initio} evaluation of these quantities, one can express them as
\begin{equation}
 \alpha_k(\iota \omega) = \langle \Psi_n | O_k | \Psi_n^{-} \rangle + \langle \Psi_n^{+} | O_k | \Psi_n \rangle
\end{equation}
with $| \Psi_n^{\pm} \rangle = \sum_{I \ne n} | \Psi_I \rangle \frac{ \langle \Psi_I | O_k | \Psi_n \rangle }{(E_I - E_n) \pm i\omega}$
which can be treated analogous to the first order wave function with respect to the ground state wave function $| \Psi_n \rangle$ due 
to the operator $D$. However, it is complicated to obtain these wave functions using sophisticated many-body methods like RCC owing 
to the presence of the imaginary factor in the denominator. Alternatively, we try to determine the ground and singly excited state 
wave functions of these systems using the following procedure. Indeed these states can be treated as a closed-shell 
configuration with a respective valence electron in the outermost orbital. We, therefore, calculate the Dirac-Fock (DF) wave function ($|\Phi_0\rangle$) for 
the closed-shell configuration first and then define the DF wave function of the ground or singly excited states of the considered systems
by appending the valence orbital ($v$) to the DF wave function of the closed-shell as $\vert \Phi_v \rangle = a_v^{\dagger}|\Phi_0\rangle$.
The exact atomic wave functions of these states can now be evaluated by considering the correlations among the electrons 
within $|\Phi_0\rangle$ which is referred as core correlation, correlations seen by the valence and core electrons of 
$|\Phi_v\rangle$ termed as valence correlation and the correlations between the core electrons with the valence 
electron $v$ named as the core-valence contributions. Using the wave operator formalism, we can write these wave functions accounting the above correlations independently as
\begin{eqnarray}
|\Psi_v \rangle &=& a_v^{\dagger} \Omega_c |\Phi_0 \rangle + \Omega_{cv} |\Phi_v \rangle + \Omega_v |\Phi_v \rangle,
\label{eqn21}
\end{eqnarray}
where $\Omega_c$, $\Omega_{cv}$ and $\Omega_v$ are known as the wave operators for the core, core-valence and valence correlations, 
respectively. 

\begin{table*}[t]
\caption{\label{range} Calculated values of effective $R^*$ range of the atom-ion interaction potentials and energy scale 
for the given atom-ion system. The results are compared with other theoretical works wherever available.}
\begin{ruledtabular}
\begin{tabular}{lcc|lcc}
 & $R^*$(in au) & $E^* \times 10^{11}$(au)  &  & $R^*$(in au) & $E^* \times 10^{11}$(au)\\
\hline
& & \\
Li-Ca$^+$    & 1336 &0.09& K-Ca$^+$  & 3231 & 1.88\\
Li-Sr$^+$    & 1393 &0.11& K-Sr$^+$  & 3779 & 3.52 \\
Li-Ba$^+$    & 1412 &0.12& K-Ba$^+$  & 3999 & 4.41 \\
Li-Ra$^+$    & 1423 &0.13& K-Ra$^+$  & 4193 & 5.33 \\
Na-Ca$^+$    & 2079 &0.57& Rb-Ca$^+$ & 3991 & 3.95 \\
             & 2081~\cite{idiaszek,doerk} & - &   & 3989~\cite{idiaszek,doerk} & -\\
Na-Sr$^+$    & 2324 &0.89& Rb-Sr$^+$ & 5042      & 10.03\\
Na-Ba$^+$    & 2412 &1.04& Rb-Ba$^+$ & 5545      & 14.6\\
Na-Ra$^+$    & 2486 &1.17&           & 5544~\cite{idiaszek,doerk} & -\\
             &      &                & Rb-Ra$^+$ & 6042           & 20.57 \\
\end{tabular} 
\end{ruledtabular}
\end{table*}
With the above prescription, the square of the matrix element of $O_k$ from Eq. (\ref{poleq}) can be expressed as
\begin{eqnarray}
\left<\Psi_v|O_k|\Psi_I\right>^2 &=& \left<\Psi_v|O_k|\Psi_I\right>\left<\Psi_I|O_k|\Psi_v\right> \nonumber \\
&=& \left<\Phi_0| \Omega_c^{\dagger} O_k [\Omega_I \Omega_I^{\dagger} + \Omega_{cI} 
\Omega_{cI}^{\dagger}] O_k  \Omega_c |\Phi_0\right>  \nonumber \\
&+& \left<\Phi_v| \Omega_v^{\dagger} O_k [\Omega_{cI} \Omega_{cI}^{\dagger} + \Omega_I \Omega_I^{\dagger}] O_k \Omega_v |\Phi_v\right> \nonumber \\
&+& \left<\Phi_v| \Omega_v^{\dagger} O_k \Omega_c \Omega_c^{\dagger} O_k  \Omega_v |\Phi_v\right> \nonumber \\
&+& \left<\Phi_I| \Omega_{cI}^{\dagger} O_k \Omega_c \Omega_c^{\dagger} O_k  \Omega_{cv} |\Phi_v\right>,
\label{mateq}
\end{eqnarray}
where we have used the generalized Wick's theorem to assemble different terms and assumed all the operators are in normal ordered form 
so that only the connected terms survive. For the brevity, we categorize the first term as core ($c$), the next two terms as valence 
($v$) and the last term as core-valence ($cv$) contributions, for which we can now write the total polarizability as
\begin{eqnarray}
\alpha_k  &=& \alpha_k^c + \alpha_k^v + \alpha_k^{cv},
\label{eq26}
\end{eqnarray}
for the notations $\alpha_k^c$, $\alpha_k^v$ and $\alpha_k^{cv}$ corresponding to the above mentioned three  
correlation contributions, respectively.

It is possible to evaluate dominant contributions to $\alpha_k^v$ by calculating many low-lying singly excited 
states $|\Psi_I \rangle$ of the considered systems by expressing them as
\begin{eqnarray}
\alpha_k^{v} (\iota\omega) &=& \frac{2}{(2k+1)(2J_n+1)} \nonumber \\ && \times \sum_{I \ne n}^{(')} \frac{ (E_n-E_I) |\langle \Psi_n || O_k || \Psi_I\rangle |^2}
{(E_n - E_I)^2 + \omega^2},
\end{eqnarray}
where $\langle \Psi_n || O_k || \Psi_I\rangle$ is the reduced matrix element of $O_k$ and the symbol $(')$ in the summation implies 
that only the excited states are included in the sum. In order to determine the E1 and E2 
matrix elements between the ground state wave function $|\Psi_n\rangle$ and the excited state wave function $|\Psi_I\rangle$, 
we express them in a general form as $|\Psi_v \rangle$ with a common core and for a valence orbital $v$ representing 
either $n$ or $I$, which in the Fock-space RCC formalism is defined as
 \begin{equation}
   |\Psi_v \rangle = e^T\{1+S_v\} |\Phi_v\rangle.
 \end{equation}
Here the operator $T$ and $S_v$ excite core electrons and the valence electron along with the core electrons due to
the electron correlations. We consider all possible single and double excitations with the important valence triple 
excitations in our calculations (referred as CCSD(T) method in the literature) within a sufficiently large configuration
space. From the practical limitation, we calculate as many as $|\Psi_I \rangle$ states possible for the estimation 
of their contributions to $\alpha_k^v$ and refer as main contribution ($\alpha^{vm}$). Contributions from the higher 
excited states, which are relatively small, are estimated using the following equation at the DF approximation  
\begin{eqnarray}
\alpha_k^{vt} (\iota\omega)= \langle \Psi_n| O_k | \Psi_n^{(1)} \rangle,
\end{eqnarray}
where the $ | \Psi_n^{(1)} \rangle$ is obtained by solving the following inhomogeneous equation for the effective 
Hamiltonian $H_{eff}=(H-E_n)O_k$ as
\begin{eqnarray}
 [(H-E_n)^2+\omega^2]| \Psi_n^{(1)} \rangle = - H_{eff} | \Psi_n \rangle
 \label{fsoeq}
\end{eqnarray}
and given as tail contribution ($\alpha_k^{vt}$). 

We also obtain the $\alpha_k^{cv}$ contributions using the same procedure as has been described by the above equation.
Nonetheless, the $\alpha_k^c$ contributions may not be smaller to be estimated using the DF method for which we employ
the random phase approximation (RPA) to solve for the core configuration (denoted by subscript 0) with the similar 
logic as Eq. (\ref{fsoeq}) by defining 
\begin{eqnarray}
| \Psi_0^{(1)} \rangle &=&  \sum_{\beta}^{\infty} \sum_{p,a} \Omega_{a \rightarrow p}^{(\beta, 1)} |\Phi_0\rangle \nonumber \\
    &=& \sum_{\beta=1}^{\infty} \sum_{pq,ab} { \{} \frac{[\langle pb | \frac{1}{r_{12}} | aq \rangle 
- \langle pb | \frac{1}{r_{12}} | qa \rangle] \Omega_{b \rightarrow q}^{(\beta-1,1)} } {(\epsilon_p - \epsilon_a)^2+\omega^2}  \nonumber \\ 
&& + \frac{ \Omega_{b \rightarrow q}^{{(\beta-1,1)}^{\dagger}}[\langle pq | \frac{1}{r_{12}} | ab \rangle - \langle pq | \frac{1}{r_{12}} | ba \rangle] 
}{(\epsilon_p-\epsilon_a)^2+\omega^2} { \}} \nonumber \\
 && \times (\epsilon_p -\epsilon_a) |\Phi_0\rangle,
\label{eqrpa}
\end{eqnarray} 
where $\Omega_{a \rightarrow p}^{(\beta, 1)}$ is the wave operator that excites an occupied orbital $a$ of $|\Phi_0 \rangle$ 
to a virtual orbital $p$ which alternatively refers to a singly excited state with respect to $|\Phi_0 \rangle$ with  
$\Omega_{a \rightarrow p}^{(0,1)} = \frac{ \langle p | (\epsilon_p-\epsilon_a)O_k | a \rangle} 
{(\epsilon_p - \epsilon_a)^2+\omega^2}$ for the single particle orbitals energies $\epsilon$s and the superscripts 
$\beta$ and 1 representing the number of the Coulomb ($\frac{1}{r_{12}}$) and $O_k$ operators, respectively. 
 
\section{Results and Discussion}\label{sec4}

\subsection{Calculation of lifetimes of the $np$ states}\label{life}
As a test of accuracy of our calculated principal matrix elements which are going to contribute predominantly to the 
$\alpha_1$ results of the alkaline earth ions, we estimate the lifetimes ($\tau$s) of the $np$ states using these matrix 
elements with $n$ being the principal quantum number of the ground states of the respective ions and compare them with the
experimental and other high precision calculations. These values are given in Table ~\ref{lifetime} and are estimated 
considering only the dominant E1 transition probabilities ($A$), which are evaluated (in $s^{-1}$) using the formula
\begin{equation}
 A^{E1}_{ij} = \frac{2.02613 \times 10^{15}}{\lambda^{3}͉͉} \frac{|\langle{i}\|D\|{j}\rangle|^2}{2j_{i}+1}, 
\end{equation}
where $\lambda$ is the wavelength of the transition in \r{A} and ${|\langle{i}\|D\|{j}\rangle|^2}$ is the reduced E1
matrix elements in au. Since our aim is to know the accuracies of the E1 matrix elements alone, we use the experimental
$\lambda$ values in these calculations. As can be seen from the table, the experimental results have large error
bars however our calculated values are compared with another high precision calculations \cite{safro} in Ca$^+$. The 
lifetimes of the $5p_{1/2,3/2}$ states of Sr$^+$  and the $6p_{1/2,3/2}$ states of Ba$^+$ are observed by 
Gallagher~\cite{Gallagher} using the Hanle-effect method with the optical excitations from the ground states. These
values are 7.35(0.3) $ns$ and 6.53(0.2) $ns$ for the $5p_{1/2}$ and $5p_{3/2}$ states of Sr$^+$, respectively, which
are later improved by Pinnington~\textit{et. al.}~\cite{Pinnington}. Our results are close to these values and the 
used E1 matrix elements can be used further to estimate $\alpha_1$ of Sr$^+$ within a reasonably accuracy. Similarly,
the experimental lifetimes of the $6p_{1/2,3/2}$ states of Ba$^+$ are reported as $\tau(6p_{1/2}) = 7.74(0.4) ns$ and 
$\tau(6p_{3/2}) = 6.27(0.25) ns$ \cite{Gallagher} and other theoretical values are given as $\tau(6p_{1/2}) = 7.83 ns$ 
and $\tau(6p_{3/2}) = 6.27 ns$ \cite{Tchoukova} which are in good agreement with our results suggesting that when the
corresponding E1 matrix elements are used, we will be able to achieve high accuracy $\alpha_1$ value in Ba$^+$. There
are no experimental results available for the lifetimes of the $7p_{1/2,3/2}$ states of Ra$^+$, however
our results are close agreement with another calculations by Pal~\textit{et.al.}~\cite{palR}. Therefore, the resulting 
$\alpha_1$ values in all the above discussed ions will be reliable and hence we expect to attain accurate values of the
dispersion coefficients when $\alpha_1$ values are used from our calculations.

\begin{table*}[t]
\caption{\label{t1} Individual contributions to $\alpha_1$ of Ca$^+$, Sr$^+$, Ba$^+$ and Ra$^+$ alkaline earth ions from 
the principal E1 matrix elements and other components. Our results are also compared with other calculations and experimental 
values.}
\begin{ruledtabular}
\begin{tabular}{lcc|lcc}
Contributions & $E1$ amplitude & $\alpha_1$ & Contributions & $E1$ amplitude &$\alpha_1$\\
\hline 
& & \\
& \text{Ca$^+$} & & & \text{Sr$^+$} & \\
\hline
& & \\
 $4s_{1/2} \longrightarrow 4p_{1/2}$  & 2.91 & 24.64 & $5s_{1/2}\longrightarrow 5p_{1/2}$ & 3.12 & 29.82  \\
 $4s_{1/2} \longrightarrow 5p_{1/2}$  & 0.07 & 0.05  & $5s_{1/2}\longrightarrow 6p_{1/2}$ & 0.02 & 0.01   \\
 $4s_{1/2} \longrightarrow 6p_{1/2}$  & 0.08 & 0.05  & $5s_{1/2}\longrightarrow 7p_{1/2}$ & 0.06 & 0.004  \\
 $4s_{1/2} \longrightarrow 7p_{1/2}$  & 0.06 & 0.004 & $5s_{1/2}\longrightarrow 8p_{1/2}$ & 0.05 & 0.003  \\
 $4s_{1/2} \longrightarrow 8p_{1/2}$  & 0.05 & 0.002 & $5s_{1/2}\longrightarrow 5p_{3/2}$ & 4.39 & 57.61  \\    
 $4s_{1/2} \longrightarrow 9p_{1/2}$  & 0.04 & 0.001 & $5s_{1/2}\longrightarrow 6p_{3/2}$ & 0.04 & 0.002  \\
 $4s_{1/2} \longrightarrow 4p_{3/2}$  & 4.12 & 48.86 & $5s_{1/2}\longrightarrow 7p_{3/2}$ & 0.05 & 0.003  \\
 $4s_{1/2} \longrightarrow 5p_{3/2}$  & 0.08 & 0.01  & $5s_{1/2}\longrightarrow 8p_{3/2}$ & 0.05 & 0.002  \\
 $4s_{1/2} \longrightarrow 6p_{3/2}$  & 0.10 & 0.012 &                                    &      & \\
 $4s_{1/2} \longrightarrow 7p_{3/2}$  & 0.08 & 0.01  &                                    &      & \\
 $4s_{1/2} \longrightarrow 8p_{3/2}$  & 0.07 & 0.004 &                                    &      & \\
 $4s_{1/2} \longrightarrow 9p_{3/2}$  & 0.06 & 0.003 &                                    &      & \\
 $\alpha_{c}$                     &      & 3.25                    & $\alpha_{c}$               &     &4.98       \\
 $\alpha_{tail}$                      &      & 5.51$\times 10^{-2}$    & $\alpha_{tail}$           &     & 1.96$\times 10^{-2}$ \\
 $\alpha_{vc}$                        &      &-8.85$\times 10^{-2}$    &$\alpha_{vc}$              &     &-0.19  \\
 $\alpha_{total}$(Present)            &      &76.89                    & $\alpha_{total}$(Present) &     & 92.25       \\
 $\alpha_{total}$(Other)              &      &75.88~\cite{Ivan}        & $\alpha_{total}$(Other)   &     &91.10~\cite{Ivan} \\
 $\alpha_{total}$(Other)              &      &75.49~\cite{ca-mitroy}   & $\alpha_{total}$(Other)   &     &91.3(9)~\cite{jiang} \\
 $\alpha_{total}$(Expt.)              &      &75.3(4)~\cite{edward}& $\alpha_{total}$(Expt.)   &     &93.3(9)~\cite{Bromley}\\
 \hline
 & & \\
  & {\text{Ba$^+$}} &  & &\text{Ra$^+$} & \\
\hline
& & \\
$6s_{1/2} \longrightarrow 6p_{1/2}$  & 3.36 & 40.76     & $7s_{1/2}\longrightarrow  7p_{1/2}$ & 3.28 & 36.86\\ 
$6s_{1/2} \longrightarrow 7p_{1/2}$  & 0.10 & 0.02      & $7s_{1/2}\longrightarrow  8p_{1/2}$ & 0.04 & 0.002\\
$6s_{1/2} \longrightarrow 8p_{1/2}$  & 0.11 & 0.016     & $7s_{1/2}\longrightarrow  9p_{1/2}$ & 0.09 & 0.01\\
$6s_{1/2} \longrightarrow 6p_{3/2}$  & 4.73 & 74.55     & $7s_{1/2}\longrightarrow  7p_{3/2}$ & 4.54 & 57.53\\
$6s_{1/2} \longrightarrow 7p_{3/2}$  & 0.17 & 0.04      & $7s_{1/2}\longrightarrow  8p_{3/2}$ & 0.49 & 0.03\\
$6s_{1/2} \longrightarrow 8p_{3/2}$  & 0.11 & 0.02      & $7s_{1/2}\longrightarrow  9p_{3/2}$ & 0.30 & 0.10\\
$\alpha_{c}$                          &      &9.35       & $\alpha_{c}$                        &      & 11.66 \\
$\alpha_{tail}$                      &      &1.66$\times 10^{-2}$  & $\alpha_{tail}$          &      & 0.15 \\
$\alpha_{vc}$                        &      &-0.38                 & $\alpha_{vc}$            &      &-0.74 \\
$\alpha_{total}$(Present)            &      &124.40                & $\alpha_{total}$(Present)&      &105.91 \\ 
$\alpha_{total}$(Other)              &      &123.07~\cite{Ivan}    & $\alpha_{total}$(Other)  &      & 105.37 ~\cite{Ivan} \\
$\alpha_{total}$(Other)              &      &126.2~\cite{Miadokova}& $\alpha_{total}$(Other)  &      & 106.5~\cite{safronova}\\
$\alpha_{total}$(Expt.)              &      &123.88(5)~\cite{Snow}
\end{tabular} 
\end{ruledtabular}
\end{table*}

\subsection{Calculation of $C_4$ coefficients }\label{C4}
In Table~\ref{pol11}, we present the static dipole polarizabilities of the alkali atoms that were reported by us in Ref.~\cite{sahoobindiya} and compare
with the earlier theoretical and experimental results. The details of the calculations are presented in Ref.~\cite{sahoobindiya} and we do not repeat them here again. The reported values of $\alpha_1$ are slightly different than
Ref. \cite{sahoobindiya}, since the core contributions from the DF method are replaced by the RPA values here. From the comparison
between the measured and calculated results, as shown in the table, it is clear that our static polarizabilities are in 
close agreement with the experimental and theoretical values which gives us confidence in using these values for 
the calculation of the $C_4$ coefficients as $82.1$, $81.2$, $144.9$ and $156.0$ au in the Li, Na, K. and Rb atoms respectively. Using 
these $C_4$ values, we further obtain range of potential $R^*$  and compare them with the values obtained by 
Idziaszek~\cite{idiaszek} and Deork~\cite{doerk}, as shown in Table \ref{range}. Idziaszek and Deork have applied the 
multichannel quantum defect theory to describe the range of the atom-ion systems. On comparison, we observe that our 
calculated values of range are close to the values tabulated in these references. From the table, we note that the effective length scale of the atom-ion potential is much more long ranged than the interaction between two neutral atoms. In the same table we also present 
the characteristic energies for their direct applications in the future experimental studies. 

\begin{table*}[htb!]
\caption{\label{neff}Calculated dispersion coefficients in this work and the estimated effective number of electrons 
values for the Slater-Kirkwood formula in case of alkali dimers. Results are compared with other available values whose
references are given in the square brackets.}
\begin{ruledtabular}
\begin{tabular}{lcccc|lcc}
 & $c_{6}$(Present) & $c_{6}$~\cite{babb} & $N_{eff}$(Present)  & $N_{eff}$~\cite{Koutselos2} &  & $c_{6}$(Present)& $N_{eff}$(Present)\\
\hline
& & \\
Li-Li & 1390 & 1389  & 0.77 & 0.773 &  Ca$^+$-Ca$^+$ & 562  & 1.24  \\
Na-Na & 1549 & 1556  & 0.99 & -     &  Sr$^+$-Sr$^+$ & 831  & 1.52   \\
K-K   & 3895 & 3897  & 1.10 & 1.13  &  Ba$^+$-Ba$^+$ & 1436 & 1.89 \\
Rb-Rb & 4663 & 4691  & 1.19 & 1.20  &  Ra$^+$-Ra$^+$ & 1341 & 2.64  \\
\end{tabular} 
\end{ruledtabular}
\end{table*}

\begin{figure}[t]
\includegraphics[width=8.5cm,height=8cm]{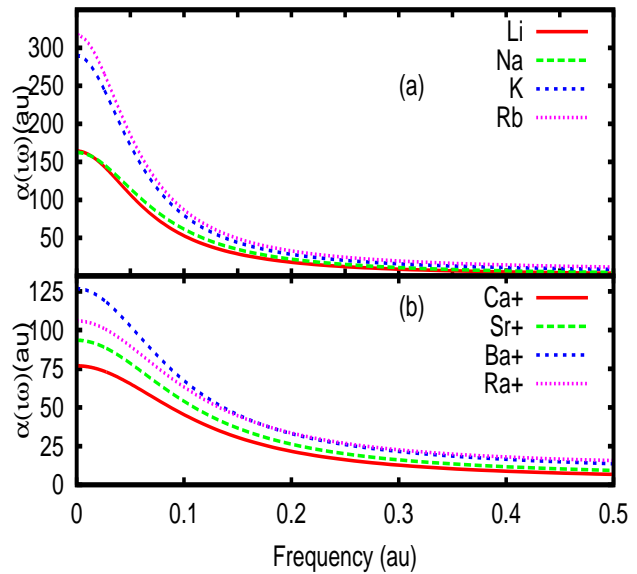}
\caption{(Color online) Comparison of the dynamic polarizabilities ($\alpha(i\omega)$) among the (a) alkali atoms and 
(b) alkaline earth ions against the frequency ($\omega$) values.}
\label{pol}     
\end{figure}

\begin{figure}[t]
\includegraphics[width=9.5cm,height=10.5cm]{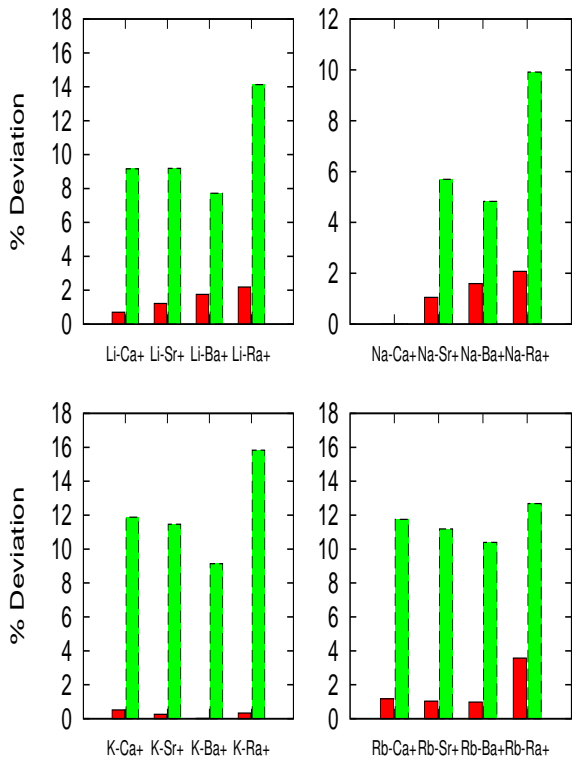}
\caption{(Color online) Percentage deviations in the $c_6^{AB}$ values obtained from this work with the results obtained using the 
Slater-Kirkwood formula (shown in the red bars) and those obtained using the approximate approach used in Ref. \cite{babb}
(shown in the green bars).}.
\label{all}     
\end{figure}

\begin{figure}[t]
\includegraphics[width=9.0cm,height=9.0cm]{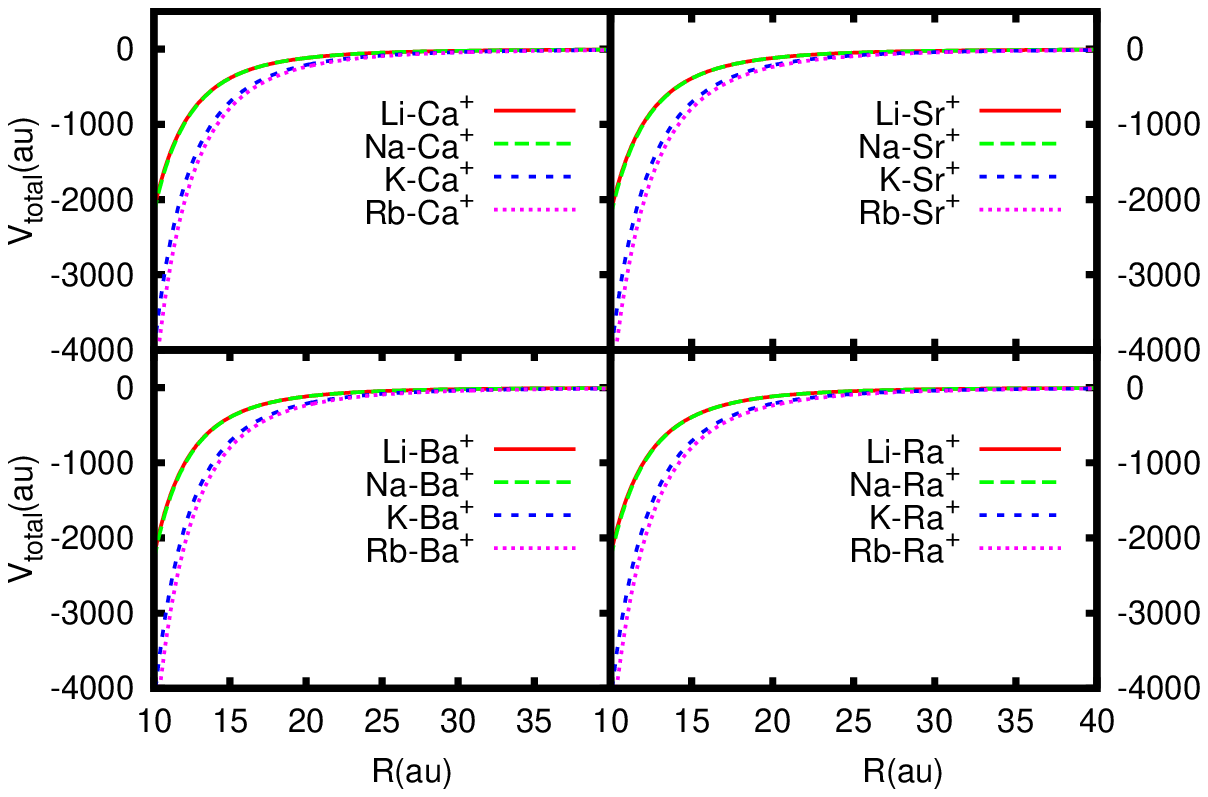}
\caption{(Color online) Net interaction potentials ($V_{total}$) between different combinations of the alkali atoms
and alkaline ions with respect to the internuclear distance ($R$). }
\label{Vtotal}     
\end{figure}

\subsection{Calculation of $C_6$ coefficients }\label{C6}

In order to obtain the $C_6$ coefficients, we first carry out systematic calculations of the quadrupole polarizabilities 
of the Li, Na, K, and Rb atoms. As given in Table~\ref{pol11}, terms $\alpha^{c}_2$ ,$\alpha^{v}_2$ and $\alpha^{vc}_2$ summarizes the contributions 
to the quadrupole polarizbailities from the core, valence and valance-core correlation terms. Here the matrix elements of 
the first five $ns-n'd_{5/2}$ transitions in each alkali atoms are included into the main term $\alpha_2^{vm}$ 
calculations, where $n$ is the principal quantum number of the ground state of the respective atom. For example in the 
Na atom, $3s$ to (3-7)$d_{5/2}$ transition E2 matrix elements are included in the main polarizability calculations. 
Moreover, for the Li atom calculation, we have included two more transitions $2s-8d_{5/2}$ and $2s-9d_{5/2}$ in the main 
polarizability calculations. In the above table, we compare our results with the predictions by other studies. For the Li
atom, accurate value of the quadrupole polarizability is obtained as 1424 au by Porsev \textit{et.al.}~\cite{qporsev} 
using the relativistic many-body calculations. Our result $1426$ au is in very good agreement with this value. Theoretical
values of the quadrupole polarizability of the Na atom were given by the group of Spelsberg~\cite{spelsberg} and 
Makarov~\cite{makarov} as $1879$ and $1902$ au respectively and are also in close agreement with our value $1895$ au. 
Group of Makarov had calculated the quadrupole polarizability of Na by using the M$\ddot{\rm{o}}$ller Plesset second-order
perturbation theory with an extended Gaussian basis. From the earlier studies, the results available to compare the 
polarizability values of the K and Rb atoms are $5000$ and $6459$ au~\cite{Marinescu} by using a model potential method 
showing only small variations from our results $4947$ and $6491$ au respectively. The comparisons, as given in 
Table~\ref{pol11}, reflects that our polarizabilities are reliable enough for the accurate determination of the 
$C_6$ results of the alkali atoms. Our numerical calculations for the $C_6$ coefficients give the values as $713$, $947$, 
$2474$ and $3245$ in au for the Li, Na, K, and Rb atoms respectively. 

\begin{table*}[htb!]
\caption{\label{c6} Comparison of the $c_6$ coefficients for the alkali atoms and alkaline earth ions with various 
methods.}
\begin{ruledtabular}
\begin{tabular}{lccccccc}
 & \multicolumn{5}{c}{Individual contributions to the $c_6$ coefficients from this work} & \multicolumn{2}{c}{Others} \\ 
 \cline{2-6} \\
& $\left|\alpha_v\right|^2$ & {$\left|\alpha_c\right|^2$} & {$\left|\alpha_{vc}\right|^2$} & {$\alpha_{\rm{c.t.}}$} & Total & (S.Kirkwood) & {$c_6$}(Ref.\cite{babb}$^a$)\\
\hline
& & \\
\;Li-Ca$^+$  & 768.4  & 1.0   & 0.0                 & 67.35  & 836.7   & 844.0   & 914.9\\
\;Li-Sr$^+$  & 898.1  & 1.5   & 0.0                 & 108.4  & 1008.1  & 1025.6  & 1106.3\\   
\;Li-Ba$^+$  & 1127.1 & 2.6   & 0.0                 & 184.8  & 1314.5  & 1355.3  & 1434.7\\ 
\;Li-Ra$^+$  & 974.3  & 3.2   & 0.0                 & 228.9  & 1206.4  & 1256.2  & 1402.9 \\
\;Na-Ca$^+$  & 826.4  & 4.6   & 2.2$\times 10^{-3}$ & 69.7   & 900.8   & 906.5   & 951.7\\
\;Na-Sr$^+$  & 964.6  & 7.4   & 4.3$\times 10^{-3}$ & 113.6  & 1085.7  & 1101.7  & 1152.4\\
\;Na-Ba$^+$  & 1205.7 & 12.4  & 7.4$\times 10^{-3}$ & 195.6  & 1413.5  & 1454.4  & 1500.7\\  
\;Na-Ra$^+$  & 1046.5 & 15.1  & 1.3$\times 10^{-2}$ & 244.3  & 1305.8  & 1357.8  & 1462.0\\
\;K-Ca$^+$   & 1231.4 & 19.6  & 1.0$\times 10^{-2}$ & 141.0  & 1392.0  & 1388.9  & 1561.9\\
\;K-Sr$^+$   & 1441.5 & 31.6  & 2.0$\times 10^{-2}$ & 207.1  & 1680.3  & 1687.4  & 1885.7 \\
\;K-Ba$^+$   & 1816.2 & 53.0  & 3.6$\times 10^{-2}$ & 328.3  & 2197.5  & 2231.7  & 2435.2\\ 
\;K-Ra$^+$   & 1563.4 & 66.1  & 6.5$\times 10^{-2}$ & 389.9  & 2019.4  & 2056.8  & 2390.4\\
\;Rb-Ca$^+$  & 1312.9 & 29.6  & 1.8$\times 10^{-2}$ & 184.8  & 1527.9  & 1517.5  & 1715.9\\
\;Rb-Sr$^+$  & 1537.3 & 48.4  & 3.8$\times 10^{-2}$ & 259.5  & 1845.2  & 1843.5  & 2071.0\\
\;Rb-Ba$^+$  & 1938.1 & 81.5  & 6.7$\times 10^{-2}$ & 394.5  & 2414.2  & 2438.4  & 2672.4\\
\;Rb-Ra$^+$  & 1667.2 & 101.6 & 0.1                 & 455.5  & 2224.5  & 2246.4  & 2625.0\\  
\end{tabular} 
\end{ruledtabular}
$^a$Note: These values do not appear explicitly in the reference, but were deduced using Eq. (\ref{aa}) quoted 
therein.
\end{table*}

\subsection{Calculations of dispersion coefficients $c_6^{AB}$}\label{c_6}
Table~\ref{c6} gives the compiled values of contributions to the total dispersion coefficients $c_6^{AB}$ between  
the alkali atoms interacting with the alkaline ions. For the determination of the dispersion coefficients, we perform 
the RCC calculations to obtain the dipole matrix elements for the evaluation of the required dipole polarizabilities of 
the Ca$^+$, Sr$^+$, Ba$^+$, and Ra$^+$ ions. There are several calculations of the ground-state polarizabilities of the 
alkaline earth ions available using different methods. Similarly, a number of precise measurements of these quantities are 
also reported in the literature. We have compared these results with the present work in Table~\ref{t1}. As can be seen 
from the table, Lim~\textit{et.al.}~\cite{Ivan} listed the static dipole polarizabilities of the considered alkaline earth 
ions which are in very close agreement with our values. Their values are predicted using the RCC calculations in the 
finite field gradient technique together with the optimized Gaussian-type basis set. However, use of a sum over states 
approach allowed us to use experimental data wherever available, which we believe that can minimize the uncertainties in the
results and hence, they are more accurate in our case. Experimental spectral analysis of the dipole polarizability value 
of the Ca$^+$ ion is observed by Edward~\cite{edward} and is in very good agreement with our calculated value.
As seen from the given table, the calculated values of these quantities by Mitroy~\textit{et.al.}{~\cite{ca-mitroy}}, 
which are evaluated by diagonalizing the semi-empirical Hamiltonian in a large dimension single electron basis, are also 
in agreement with our values. However, we would like to emphasize that our results are more accurate since in our method 
core correlations are accounted through the all order RPA. The estimate of 93.3 au~\cite{SR} for the ground state 
polarizability of Sr$^+$ ion, derived by combining the experimental data given by the group of Barklem~\cite{Bromley}
with the oscillator strength sums, has a considerable discrepancy with our present results. In contrast, our values match 
very well with the calculations of Jiang~\textit{et.al.}{~\cite{jiang}} who have used the relativistic all-order method to
calculate the polarizabilities of the Sr$^+$ ion. It would be interesting to see the validity of these results when the 
new measurement of the ground state polarizability for this ion becomes available. The polarizability value of the 
Ba$^+$ ion was calculated by Miadokova~\textit{et.al.}{~\cite{Miadokova}} using the relativistic basis set in the 
Douglas-Kroll no-pair approximation and has a 2\% discrepancy from the high precision measurements performed by Snow and 
Lundeen~\cite{Snow}. This high precision measurement was achieved by a novel technique based on the resonant excitation 
Stark ionization spectroscopy microwave technique. We also find that our results are in better agreement with the 
experimental value. There is no experimental result available for the dipole polarizability of Ra$^+$ to compare with
our result. However, Safronova \text{et. al.}~\cite{safronova} have evaluated this result using the relativistic all order
method and is in agreement with our result. Having compared all our polarizability results, we are now in the state to
justify that since our static polarizability values are very accurate, we anticipate similar accuracies for the calculated
dynamic polarizabilities using our method and can be used reliably for the evaluation of the dispersion coefficients.

In Fig.{\ref{pol}}, we plot our dynamic polarizabilities obtained for various atoms and ions along the imaginary axis as 
the functions of the frequencies. Next, we use the dynamic polarizabilities to calculate the dispersion coefficients and 
effective number of the electrons in case of the alkali atoms and alkaline earth ions, which are presented in 
Table~\ref{neff}. The purpose of calculating and presenting these values is to verify the validity of the results 
reported using the Slater-Kirkwood formula as given in Eq.(\ref{slater}) and using the approach that was followed
by Dereviako~\textit{et.al.}~\cite{babb}.

In practice, a number of methods have been employed for the calculations of the dispersion $c_6^{\rm{AB}}$ coefficients 
for the hetero-nuclear dimers. Dalgarno~\textit{et.al.}~\cite{dalgarno} had followed a procedure to reduce the two central 
molecular problem to one central atomic problem at the larger separation distances. Bishop and coworkers~\cite{Bishop} 
computed the $c_6^{\rm{AB}}$ coefficients by approximating the integral given by Eq. (\ref{eqc6}) using the Gaussian 
quadrature technique. In this work, we evaluate the dispersion coefficients $c_6^{AB}$ using three different methods: 
(i) the exact formula of Eq. (\ref{eqc6}), (ii) Slater-Kirkwood formula given by Eq. (\ref{slater}), and (iii) using an 
approximated approach as has been used in Ref.{\cite{babb}} (see Eq. (\ref{aa})) to make a comparative analysis among
the results obtained from all these approaches. In the case of the exact method, we use the Gaussian quadrature method to 
integrate over the dynamic polarizabilities using the exponential grids. We justify the use of the exponential grids from
the fact that maximum contributions to the integrand given in Eq. (\ref{eqc6}) come from the polarizability values in the 
vicinity of zero frequency (as shown in Fig.~\ref{pol}) and gradually their contributions falls down. In Table~\ref{c6}, 
we present details of the calculated values of the dispersion coefficients for the interactions between the alkali atoms 
and the alkaline ions along with their breakdown from the individual contributions. From the table, it can be inferred 
that the contribution to the total potential increases as the alkali atoms get bigger in size (i.e. from the Li to Rb 
sequence), since the polarizability values also increase in the same order. However, we notice that a steady increase in 
$c_6$ values do not occur with respect to the atomic sizes for the ions (i.e. from Ca$^+$ to Ra$^+$). This might seem 
to be counterintuitive but it is owing to the fact that the polarizability of Ba$^+$ is larger than that for Ra$^+$,
as given in Table~\ref{t1}. So 
it follows a different trend in the $c_6^{AB}$ coefficients; decreases in magnitude for the interactions of the alkali 
atoms with Ra$^+$ and increases with Ba$^+$. The dispersion coefficients for the atom-ion systems obtained using the
Slater-Kirkwood formula are listed in column VI of Table~\ref{c6}. In an alternative approach, we also carried out the 
calculations using the approximated formula given in Eq. (\ref{aa}) and the obtained values are listed in column VII of 
the same table. Comparison of deviations in percentage from both the approaches are shown in Fig. (\ref{all}) for all 
combinations of the alkali atoms-alkaline ions in the form of the histograms. It is apparent from this plot that the 
Slater-Kirkwood formula shows better agreement with our results as compared to the approximated approach of 
Derevianko~\textit{et. al.}.

In Fig.~\ref{Vtotal}, comparison between the total interaction potential ($V_{total}$) for the undertaken different 
combinations of Li, Na, K and Rb atoms with Ca$^+$, Sr$^+$, Ba$^+$ and Ra$^+$ ions is shown as function of internuclear 
distance $R$ by adding both the induction and dispersion parts. Interactions of each
alkaline ion (Ca$^+$, Sr$^+$, Ba$^+$, Ra$^+$) is represented in solid red line for the Li, long dashed green line for the Na, 
short dashed blue line for the K and dotted pink line for the Rb atoms, respectively. It should be noted that our results for these 
potentials will be valid in the approximation only when the structures of the colliding atom and ion do not undergo   
internal changes.

\section{Conclusion}
In this work, we have deduced the behavior of the potential curves with respect to the internuclear distances for the
alkali atoms correlating with the alkaline-earth ions. The accurate values of the dipole polarizabilities for the alkali 
atoms and the alkaline earth ions and the quadrupole polarizabilities for the alkali atoms have been investigated using the 
relativistic coupled-cluster method. Thereafter, evaluation of the dispersion coefficients have been done by integrating 
the atom-ion dynamic electric polarizabilities product at the imaginary frequencies. The calculated values of the 
induction coefficients in the form of range of potentials are expected to be very useful to set the actual positions of 
the bound states and magnetic fields of the Feshbach resonances for these atom-ion correlated systems. 
The presented data will also be of immense interest for designing better atomic clocks, quantum information processing 
and quantifying molecular potentials for the ultracold collision studies.

\section*{Acknowledgement}
The work is supported by CSIR grant no. 03(1268)/13/EMR-II, India and UGC-BSR grant no. F.7-273/2009/BSR. Computations were carried 
out using 3TFLOPHPC Cluster at Physical Research Laboratory, Ahmedabad. The authors would like to thank
R.$\rm{C}\hat{o}\rm{t}\acute{e}$ and S. Banerjee for fruitful discussions.


\end{document}